\documentclass[amsmath, aps,prc,twocolumn,groupedaddress, showkeys, showpacs]{revtex4}
\usepackage{graphicx}
\usepackage{epsfig}
\begin{document}

\title{Validity of the Wigner-Seitz approximation in neutron star crust}

\author{N. Chamel}
\affiliation{Institut d'Astronomie et d'Astrophysique, Universit\'e
Libre de Bruxelles, CP226, Boulevard du Triomphe, 1050 Brussels, Belgium}
\author{S. Naimi}
\affiliation{Institut de Physique Nucl\'eaire, Universit\'e Paris-Sud,
IN$_2$P$_3$-CNRS, F-91406 Orsay Cedex, France }
\author{E. Khan}
\affiliation{Institut de Physique Nucl\'eaire, Universit\'e Paris-Sud,
IN$_2$P$_3$-CNRS, F-91406 Orsay Cedex, France }
\author{J. Margueron}
\affiliation{Institut de Physique Nucl\'eaire, Universit\'e Paris-Sud,
IN$_2$P$_3$-CNRS, F-91406 Orsay Cedex, France }

\date{\today}

\begin{abstract}
Since the seminal work of Negele and Vautherin, the Wigner-Seitz approximation 
has been widely applied to study the inner crust of neutron stars formed of nuclear 
clusters immersed in a neutron sea. In this article, the validity of this approximation 
is discussed in the framework of
the band theory of solids. For a typical cell of $^{200}$Zr, present in the
external layers of the inner crust, it is shown that the ground state
properties of the neutron gas are rather well reproduced by the Wigner-Seitz
approximation, while its dynamical properties depend on the energy scale of
the process of interest or on the temperature. It is concluded that the Wigner-Seitz
approximation is well suited for describing the inner crust of young neutron
stars and the collapsing core of massive stars during supernovae explosions.
However the band theory is required for low temperature transport properties as, 
for instance, the effective neutron mass giving rise to 
entrainment effects.
\end{abstract}

\pacs{97.60.Jd,26.60.+c,21.10.Ma,21.60.Jz,71.15.Ap,71.18.+y,21.10.-k}
\keywords{neutron star crust, band theory, Wigner-Seitz approximation, Hartree-Fock}
\maketitle

In the standard model of neutron stars~\cite{baym71}, the crust is believed
to be formed of nuclear clusters in a body centered cubic lattice stabilized
by Coulomb forces and considered infinite and pure (made of only one type of
nuclei at a given density). In the inner crust, at densities above $\sim$
$4.10^{11}$ g.cm$^{-3}$ and below $\sim$ $10^{14}$ g.cm$^{-3}$, the
``neutron drip'' regime is reached and the clusters are surrounded by a
neutron fluid.  A formal comparison can be made with electrons in ordinary
solids present on earth: part of the neutrons participate to the nuclear
clusters which form the lattice (equivalent to electrons bounded to atoms)
while part of the neutrons are delocalized over the whole crystal
(equivalent to valence electrons). As a consequence, the band theory of
solids developed in condensed matter~\cite{kittel} can be applied to describe
the crust of neutron star. But due to the highly specific numerical issues
of band theory, nuclear physicists have preferred to use an approximation due
to Wigner and Seitz (W-S)~\cite{WS33, WS34}, where the crust is divided into
independent and spherical cells. Since the work of Negele and
Vautherin~\cite{ne73}, the W-S approximation 
has been used to predict the structure of the crust, the pairing properties,
the thermal effects, or the low lying energy excitation 
spectrum~\cite{bonche81, sandu04,sandu04b,khan05,vigez05,baldo05,baldo06}. Only
recently, band theory calculations have been carried out in order to study
the hydrodynamical properties of the neutron fluid
 and in particular the neutron effective
mass giving rise to entrainment effects~\cite{carter05,
chamel05,chamel06}, although these calculations are not yet self-consistent.
While the W-S approximation is well justified below the ``neutron drip''
regime, its validity beyond remains to be assessed. 

In this article, we investigate the limitations of the W-S approximation in
the $\rho \sim$ 7.10$^{11}$ g.cm$^{-3}$ density layer of the inner crust,
composed of a crystal of zirconium like clusters~\cite{ne73} surrounded by
the neutron gas. In Sect.~\ref{sec1}, before discussing the W-S approximation, we
briefly review the band theory of solids. Then we compare in Sect.~\ref{sec2} the
results of the band theory with those of the W-S approximation for the
single particle wavefunctions and energy spectra. Consequences for the
properties of the neutron gas are discussed.

\section{Microscopic quantum description of neutron star inner crust}
\label{sec1}

An accurate description of the inner crust, assuming that it is a perfect 
crystal, should rely on the band theory of solids~\cite{kittel}.  In this section, we briefly review
this theory in the context of neutron star crust and discuss the W-S approximation in this framework.

\subsection{Band theory of solids}

According to the Floquet-Bloch theorem, the single particle quantum states 
are given by modulated plane waves
\begin{equation}
\label{bloch_states}
\varphi_{\alpha\pmb{k}}(\pmb{r}) = u_{\alpha\pmb{k}}(\pmb{r}) e^{ {\rm i} 
\pmb{k}\cdot\pmb{r}} \, ,
\end{equation}
where the functions $u_{\alpha\pmb{k}}(\pmb{r})$ have the full periodicity
of the lattice.  Each single particle quantum state is thus labeled by a
discrete index $\alpha$ and by a wave vector $\pmb{k}$. The energy spectrum
is therefore formed of a series of sheets or ``bands'' in $\pmb{k}$-space. 

The Bloch states (\ref{bloch_states}) are completely determined by the
knowledge of the functions $u_{\alpha\pmb{k}}(\pmb{r})$ inside a W-S cell of
the lattice, whose shape is imposed by the symmetry of the crystal. The
cell, centered around one nuclear cluster, is electrically neutral and
therefore contains as many electrons as protons.  The effects of the ion
lattice on the electrons, which give rise to complicated band structures in
ordinary terrestrial matter are negligible in the inner crust of a neutron star due to
the very high densities~\cite{pethick95}. Nevertheless, the neutron band
effects due to nuclear inhomogeneities cannot be ignored.

In the present study, we consider the outermost layers of the inner crust
where pairing effects are negligible~\cite{baldo05b}. In the Hartree-Fock
approximation with Skyrme forces which we shall consider in the following,
the occupied nucleon single particle wave functions are obtained by solving
the self-consistent equations ($q=n,p$ for neutrons and protons
respectively)
\begin{equation}
\label{HF} 
h^{(q)}_0 \varphi^{(q)}_{\alpha\pmb{k}}(\pmb{r}) = 
\varepsilon^{(q)}_{\alpha\pmb{k}}
\, \varphi^{(q)}_{\alpha\pmb{k}}(\pmb{r})
\end{equation}
where the single particle Hamiltonian is defined by 
\begin{equation}
\label{h0}
h^{(q)}_0  \equiv -\pmb{\nabla}\cdot \frac{\hbar^2}{2 m^{\oplus}_q(\pmb{r})} 
\pmb{\nabla} + U_q(\pmb{r}) -{\rm i} \pmb{W_q}(\pmb{r})\cdot 
\pmb{\nabla}\times\pmb{\sigma} \, ,
\end{equation}
the effective masses $m^{\oplus}_q(\pmb{r})$, mean fields $U_q(\pmb{r})$ and
spin-orbit terms $\pmb{W_q}(\pmb{r})$ being functionals of the single
particle wave functions. These equations have to be solved inside the W-S
cell with the boundary conditions imposed by the Floquet-Bloch theorem
\begin{equation}
\label{bloch_boundary}
\varphi^{(q)}_{\alpha\pmb{k}}(\pmb{r}+\pmb{T}) = e^{ {\rm i} \pmb{k}\cdot\pmb{T}}
\varphi^{(q)}_{\alpha\pmb{k}}(\pmb{r}) \, ,
\end{equation}
where $\pmb{T}$ is any lattice vector. This means in particular that the
wavefunction between two opposite faces of the cell has a phase shift 
$e^{{\rm i} \pmb{k}\cdot\pmb{T}}$ where $\pmb{T}$ is the corresponding
lattice vector. The single particle energies are periodic in the reciprocal
lattice whose vectors $\pmb{K}$ satisfy $\pmb{K}\cdot\pmb{T}=2\pi n$ (where
$n$ is any integer)
\begin{equation}
\label{reciprocal_periodicity}
\varepsilon^{(q)}_{\alpha,\pmb{k}+\pmb{K}}=\varepsilon^{(q)}_{\alpha\pmb{k}} \, .
\end{equation}
Consequently only the values of $\pmb{k}$ inside the first
Brillouin zone (\textit{i.e.} W-S cell of the reciprocal lattice) are
relevant.

Equivalently, equations (\ref{HF}) can be written directly for the
$u^{(q)}_{\alpha\pmb{k}}(\pmb{r})$ functions in the decomposition
(\ref{bloch_states}) which leads to
\begin{equation}
\label{red_HF}
(h^{(q)}_0 + h^{(q)}_{\pmb{k}} ) u^{(q)}_{\alpha\pmb{k}} (\pmb{r}) = 
\varepsilon^{(q)}_{\alpha\pmb{k}}\,  u^{(q)}_{\alpha\pmb{k}}(\pmb{r})
\end{equation}
where the $\pmb{k}$-dependent Hamiltonian $h^{(q)}_{\pmb{k}}$ is defined by 
\begin{equation}
\label{hk}
h^{(q)}_{\pmb{k}} \equiv \frac{\hbar^2 k^2}{2 m^{\oplus}_q(\pmb{r})} + 
\pmb{v_q}\cdot\hbar\pmb{k} \, ,
\end{equation}
and the velocity operator $\pmb{v_q}$ is defined by the commutator
\begin{equation}
\label{velocity_def}
\pmb{v_q} \equiv  \frac{1}{{\rm i} \hbar}[\pmb{r}, h_0^{(q)}]  \, .
\end{equation}

The band theory takes into account all the symmetries of the system. 
However equations (\ref{HF}) with the boundary conditions
(\ref{bloch_boundary}) are numerically very complicated to solve. The
approximation introduced a long time ago by Wigner and Seitz in the study of
metallic sodium~\cite{WS33,WS34} has been widely applied in the context of
neutron star crust, as described below.

\subsection{Wigner-Seitz approximation}
\label{WS_app}

The spherical W-S approximation is a computationally very efficient method 
with the advantage of reducing the 3D partial differential Eqs.~(\ref{HF}) 
to ordinary differential radial equations. This approximation is twofold.
First of all, the Hamiltonian $h^{(q)}_{\pmb{k}}$ in equation (\ref{red_HF})
is neglected. Consequently the wave functions and the energies are
independent of $\pmb{k}$ and approximated by the solutions at $\pmb{k}=0$. 
Only the band index $\alpha$ remains.  Secondly, the W-S polyhedron is
replaced by a sphere of equal volume. The equations are then usually solved
with the Dirichlet-Neumann mixed boundary conditions which yield a 
nearly constant neutron density outside the cluster.

The W-S approximation turns out to be very good if the boundary
conditions play a minor role. For instance, bound states whose associated
wave functions are vanishingly small outside the clusters are very well
treated provided that the spatial extent of these states is smaller than the
lattice spacing. This condition is fulfilled almost everywhere in the crust
except in the bottom layers where the clusters nearly touch. The aim of this
paper is to investigate the validity of the W-S approximation for the
outermost layers of the inner crust where the bound neutron states are not
altered by the boundary conditions.

Let us emphasize that in the W-S approximation, the nuclear clusters are 
supposed to be spherical while in the full band theory, no 
assumption is made about their shape. For the low
densities of interest in this study, the nuclear clusters can still be
considered as spherical. It should mentioned that in a recent development
of the W-S approximation~\cite{magier03}, the W-S cell is replaced by a cube
with strictly periodic boundary conditions. Possible deformations of the
nuclear clusters are thus included but at the price of unphysical boundary
conditions because the W-S cell of the body centered cubic lattice is a truncated 
octahedron and not a cube (the cube being the W-S cell of the simple cubic
lattice). This is the reason why we still consider the spherical W-S
approximation closer to the physical situation than the cubic one at low
density.

\section{Comparison between the band theory and the W-S approximation}
\label{sec2}

The comparison between the band theory and the W-S approximation gives an
estimate of the contribution (\ref{hk}) of the $k$-dependent Hamiltonian
$h_{\pmb{k}}$ and incidentally on the effects of the boundary conditions. We
have considered the shallow layers of the crust at the average baryon density 
$\rho\sim 7\times 10^{11}$ g/cm$^3$ formed of a crystal made of zirconium 
like clusters ($Z=40$) with $160$ neutrons (bound and unbound) per lattice site\cite{ne73}. 
In the following we shall refer to such cluster as $^{200}$Zr. 
Under such conditions, only unbound neutrons are sensitive to the boundary conditions 
and the nuclear clusters are spherical.

\begin{figure}
\begin{center}
\epsfig{file=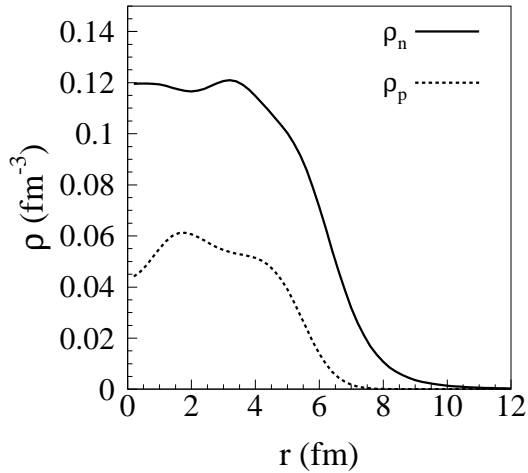,width=8.5cm}
\end{center}
\caption{Neutron (full lines) and proton (dotted lines) density
distributions for $^{200}$Zr in the W-S cell, obtained in the W-S approximation.}
\label{fig:dens}
\end{figure}

For the comparison, we first solve the self-consistent Hartree-Fock
equations in coordinate space, considering the W-S approximation. The
effective force used is the same as in 
references~\cite{sandu04,sandu04b,khan05}, namely the Skyrme interaction 
SLy4~\cite{chabanat98}.  In Fig.~\ref{fig:dens}, we show the neutron and proton
densities calculated in the spherical W-S cell with $^{200}$Zr. The size of
the box is $R_{\rm cell}=49.2$ fm. The cell exhibits a very large and
diffuse neutron skin, which is typical of those systems~\cite{ne73}. A small
but non-zero neutron density is present at
large radius generating a non-zero mean field potential. Asymptotically,
this potential is equal to -0.05 MeV.  All the states with energy larger
than -0.05 MeV are therefore unbound or ``free''. We found that among the
160 neutrons per lattice site, 70 are unbound.

The effective mass $m^{\oplus}_n(\pmb{r})$ and the mean field potential
$U_n(\pmb{r})$ obtained for the spherical cell are used to construct an
effective Schr\"odinger equation for band theory calculations. As the
spin-orbit splitting is weak for most of the states (see Fig.~ 
\ref{fig:spectrum}), we set the spin-orbit potential $\pmb{W_n}(\pmb{r})$ to
zero. 
In order to study the effects of the boundary conditions, 
the Schr\"odinger equation is solved with no further iterations by
imposing the Bloch boundary conditions~(\ref{bloch_boundary}) and using the
Linearized Augmented Plane Wave method (see~\cite{chamel06} for details). 
The coordinate space is divided in two regions: a spherical region around
the cluster plus an interstitial region.  In the latter region, the wave
functions are expanded on a plane wave basis in order to fulfill the Bloch
boundary conditions. The lattice spacing is determined by requiring that the
volume of the W-S sphere is equal to the volume of the exact W-S cell of the
crystal, assumed to be a body centered cubic lattice\cite{baym71}.

In the following, we compare the single particle wave functions and energy spectra
of the unbound neutrons.

\subsection{Single particle wavefunctions}

\begin{figure}
\begin{center}
\epsfig{file=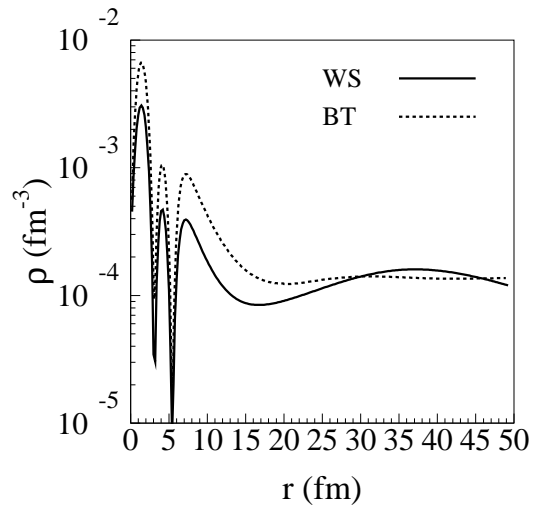,width=8.5cm}
\end{center}
\caption{Unbound neutron density calculated with the W-S approximation (WS, full
lines) and the full band theory (BT, dotted lines).}
\label{fig:densf}
\end{figure}

As already discussed in Sect.~\ref{WS_app}, the wave functions of the
bound states are nearly independent of the boundary conditions. As a
consequence, we expect that the band theory and the W-S approximation
provide identical bound states. Since the unbound states are orthogonal to
the bound states, the W-S approximation and the band theory are expected to 
yield similar unbound wavefunctions inside the nuclear clusters.  This is
confirmed by the calculation of the density distribution of the unbound 
neutrons (whose single particle energies exceed $-0.05$ MeV as discussed 
previously) shown on Fig.~\ref{fig:densf}, obtained with the band theory 
and the W-S approximation. For the comparison, the density $\rho(\pmb{r})$
obtained from the band theory, has been averaged over the solid angle around
one lattice site as follows 
\begin{equation}
\label{rho_mean}
\rho(r)=\int \frac{{\rm d}\Omega}{4\pi} \rho(\pmb{r}) \, ,
\end{equation}
where $r$ is the radial distance from the lattice site.  Similar density
oscillations are obtained in both calculations in the vicinity of the 
nuclear cluster for $r<10$ fm as expected. Qualitative differences in the
neutron density are however observed in the interstitial regions outside
clusters due to different boundary conditions. The unbound neutron density
distribution is nearly flat in the band theory while it is more fluctuating
in the W-S approximation. The Bloch wavefunctions outside the clusters are
similar to plane waves (thus giving a constant density) which cannot be
properly described in the W-S approximation owing to the expansion of the
wavefunctions into only a few spherical harmonics.  An analysis of the
contribution of each single particle wave function to the unbound neutron
density in the W-S approximation reveals that the oscillations at small
radius are mainly coming from p-states, such as 3p$_{1/2}$ or 3p$_{3/2}$
whereas at larger radii ($r>20$ fm) only a few larger $\ell$ states, mainly
d-f-g-h states, are contributing to the free neutron density.

\begin{figure}
\begin{center}
\epsfig{file=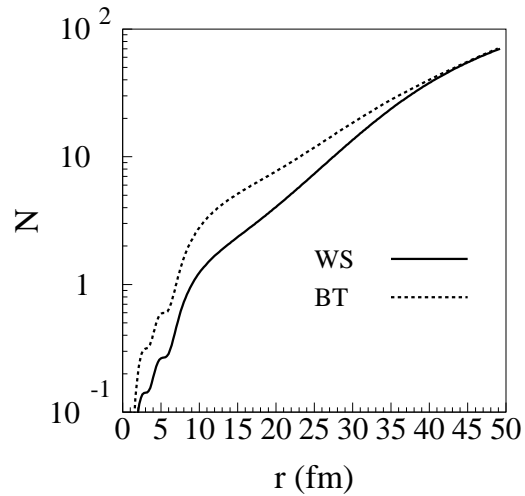,width=8.5cm}
\end{center}
\caption{Integrated unbound neutron number (see text) calculated with the 
W-S approximation (WS, full lines) and the band theory (BT, dotted lines) }
\label{fig:densfint}
\end{figure}

As a result, the W-S approximation predicts a different number of neutrons
outside the cluster than in the band theory. Since the total number of free
neutrons should be the same in both calculations, the difference in the
density profile at large radius implies a larger difference in magnitude at
small radius. This is more clearly seen on Fig.~\ref{fig:densfint}, by
plotting the integrated number $N(r)$ of free neutrons inside the W-S cell
at radius r, defined by
\begin{equation}
N(r)=4\pi \int_0^r r'^2 \rho(r') {\rm d} r'\, ,
\end{equation}
$\rho(r)$ being the local density of \emph{unbound} neutrons. The figure
shows that the W-S approximation underestimates the number $N_{\rm in}=N(R)$
of free neutrons inside the region of radius $R$ around the cluster and 
consequently overestimates the number $N_{\rm out}=N(R_{\rm cell})-N(R)$ of 
free neutrons outside.  For both calculations, the number of free neutrons
in the cell is $N(R_{\rm cell})=70$. Quantitatively taking $R=15$ fm, the
difference between the two calculations is about $\Delta N=|\Delta N_{\rm
in}|=|\Delta N_{\rm out}|=3$ which is rather small. 

This first comparison shows that the single particle wave functions of
unbound neutrons are qualitatively well reproduced by the W-S approximation
inside the nuclear clusters.  The main differences in the wavefunctions
between the two calculations are found in the interstitial region due to the
different boundary conditions. However this has a rather small effect on the
ground state properties of the neutron gas, like the neutron density 
distribution.  More generally the W-S approximation can be expected to be a
good approximation to the full band theory for evaluating the matrix
elements of any operator taking vanishing values outside the cluster region.

\subsection{Single particle energy spectrum}

\begin{figure}
\vskip 1cm
\begin{center}
\epsfig{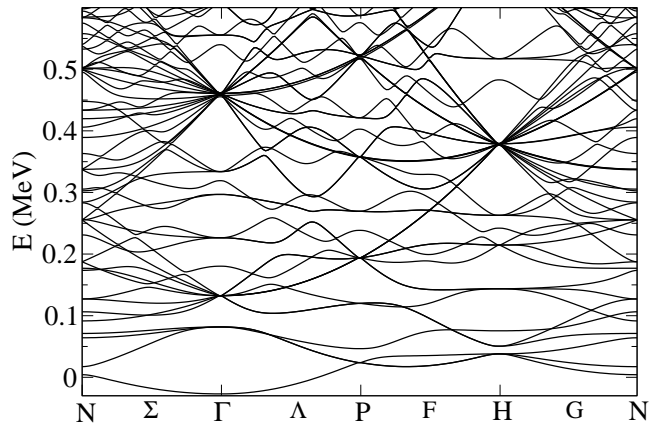}
\end{center}
\caption{Unbound single particle energy spectrum as obtained in the band 
theory vs the Bloch wave vector $\pmb{k}$, along high symmetry lines 
in the first Brillouin zone using standard notations~\cite{koster57}.
}
\label{fig:band_spectrum}
\end{figure}

\begin{figure}
\begin{center}
\epsfig{file=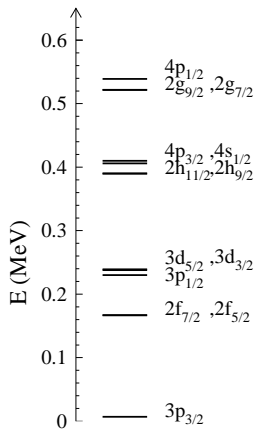,width=6cm}
\end{center}
\caption{Unbound single particle energy spectrum obtained in the W-S 
approximation.}
\label{fig:spectrum}
\end{figure}

Figs.~\ref{fig:band_spectrum} and \ref{fig:spectrum} show the energy
spectrum of the unbound neutrons obtained in the band theory and in the W-S 
approximation, respectively. It should be noted from
Fig.~\ref{fig:spectrum} that the spin-orbit splitting is very weak for d,
f, g and h states (as predicted by Negele\&Vautherin~\cite{ne73}) but not
for p states. This is due to the fact that the spin-orbit splitting is
proportional to the convolution of the density gradient together with the
wave functions. The density gradient is localized in the central cluster
while the d, f, g and h states are mostly in the external region.  For those
states, the convolution leads to a weak splitting.

Since in band theory, the energies depend also on the wavevector $\pmb{k}$,
only the energy bands along some specific symmetry directions in
$\pmb{k}$-space are displayed. The energy spectrum obtained with the W-S
approximation is comparable to the one obtained in the band theory for the
symmetry point $\Gamma$ corresponding to the center $\pmb{k}=0$ of the first
Brillouin zone (W-S cell of the reciprocal lattice). The correspondence is
not exact and the differences come from the spherical approximation. As a result,
the W-S approximation predicts less states but with larger degeneracies than
the band theory at the symmetry point $\Gamma$. The figures show clearly
another important difference between the band theory and the W-S
approximation: in the former case the energy spectrum is continuous while in
the latter case it is discrete. 

A relevant quantity to compare the energy spectra is the level density,
which plays a pivotal role when calculating dynamical processes. It is
defined by 
\begin{equation}
\label{level_density}
g(E)= V_{\rm cell} \sum_\alpha \int_{\rm BZ} \frac{{\rm d}^3\pmb{k}}{(2\pi)^3} 
\delta(E-\varepsilon_{\alpha\pmb{k}})
\end{equation}
where the integral is taken over the first Brillouin zone (BZ). 

Using the $\delta$ function to integrate out one of the variables, the
level density becomes
\begin{equation}
\label{level_density2}
g(E)=\frac{V_{\rm cell}}{(2\pi)^3} \sum_\alpha \int \frac{{\rm d} {\cal S}(E)}
{|\nabla_{\pmb{k}}\varepsilon_{\alpha\pmb{k}}|} \, ,
\end{equation}
where the integral is taken over the surface of constant energy 
$\varepsilon_{\alpha\pmb{k}}=E$ in $\pmb{k}$-space. Expression (\ref{level_density2}) 
shows that the level density is a probe of the topology 
of constant energy surfaces in $\pmb{k}$-space. 
We have extracted the level density from band theory by using the 
Gilat-Raubenheimer method as in reference~\cite{chamel05,chamel06}.

In the W-S approximation the level density reduces to a discrete sum
\begin{equation}
\label{level_density_WS}
g_{\rm WS}(E)= \sum_{n j \ell} (2 j +1) \delta(E-\varepsilon_{n j \ell}) 
\, .
\end{equation}

\begin{figure}
\begin{center}
\epsfig{file=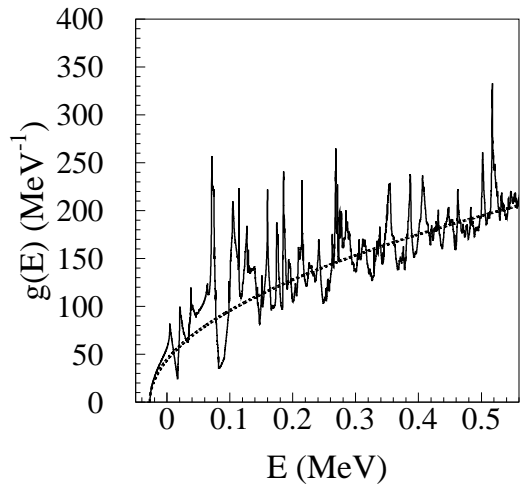,width=8.5cm}
\end{center}
\caption{Level density for neutron unbound states calculated with the 
band theory (solid line) compared with the prediction of the free Fermi gas
(dashed line). 
The energy resolution is of the order of keV.}
\label{fig:states}
\end{figure}

\begin{figure}
\begin{center}
\epsfig{file=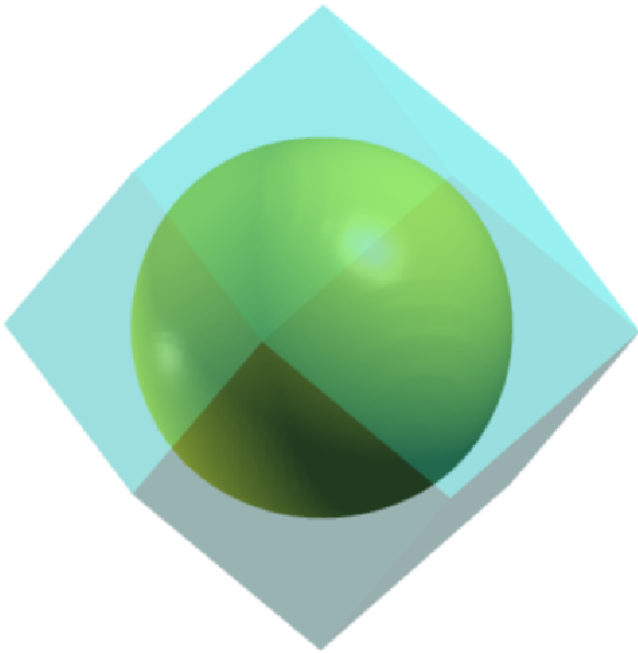,width=8.5cm}
\end{center}
\caption{(Color online)Constant neutron energy surface of a body centered cubic lattice of $^{200}$Zr inside the first Brillouin zone for $E=-0.0015 $MeV.}
\label{fig:const_surf1}
\end{figure}

\begin{figure}
\begin{center}
\epsfig{file=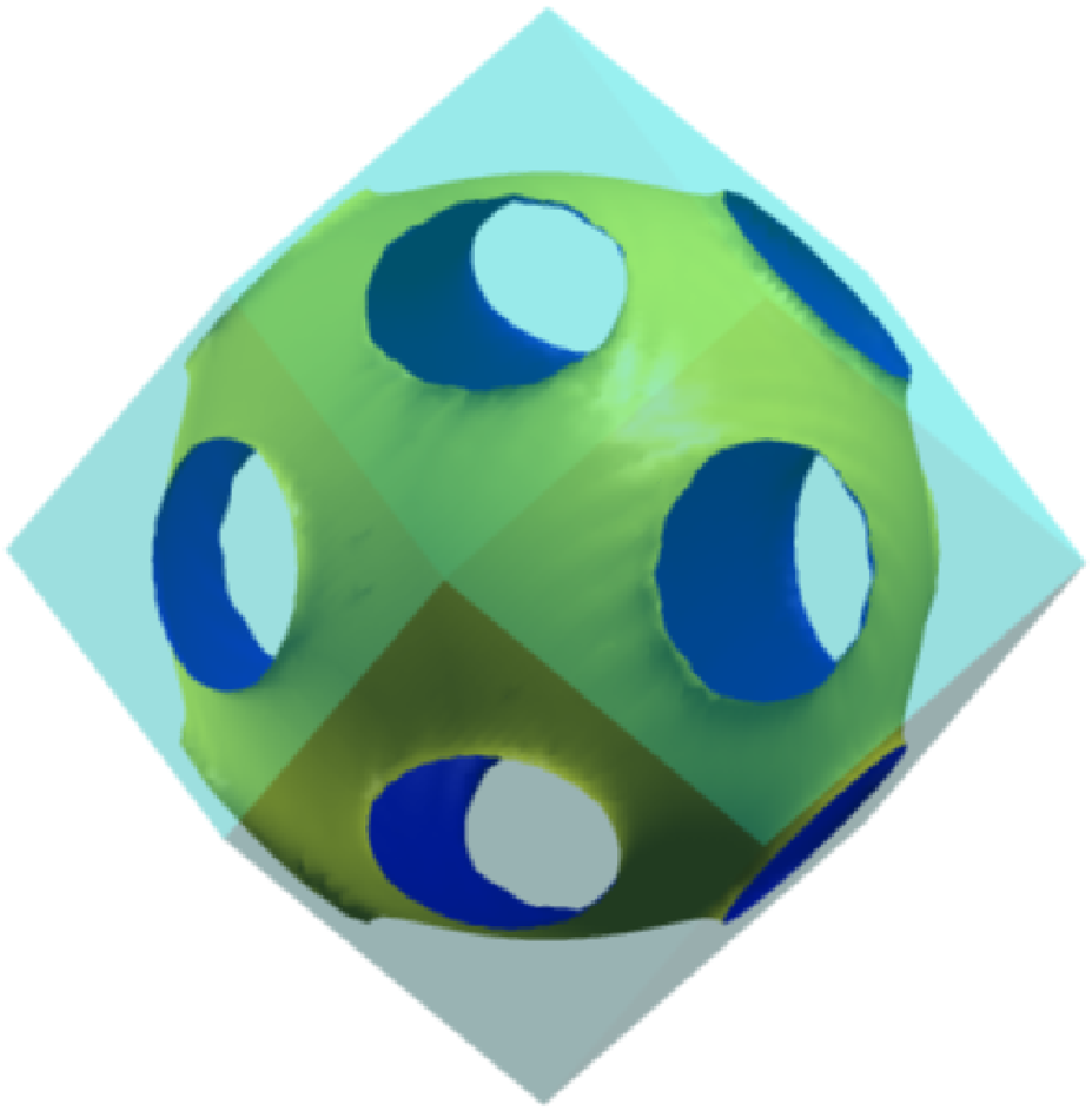,width=8.5cm}
\end{center}
\caption{(Color online)Constant neutron energy surface of a body centered cubic lattice of $^{200}$Zr inside the first Brillouin zone for $E=0.0085 $MeV.}
\label{fig:const_surf2}
\end{figure}

In Fig.~\ref{fig:states}, we show the level density predicted by the band 
theory for the unbound single particle levels.  As expected the figure shows
that the energy spectrum in the band theory is continuous and has a complex
fine structure.  The spectrum exhibits a quasi band gap of about 30 keV
slightly around 90 keV.  This is in sharp contrast with the W-S
approximation for which the energy levels are discrete and separated by
about 100 keV.  In other words, the W-S approximation overestimates the
neutron shell effects. The global energy dependence of the level density
follows the behavior of the free Fermi gas,
\begin{equation}
\label{level_density_fermi_gas}
g(E)\simeq \frac{V_{\rm cell}}{2\pi^2} \biggl(\frac{2m}{\hbar^2}\biggr)^{3/2}\sqrt{E-E_v}\, ,
\end{equation}
where $E_v\simeq-0.031$ MeV is the energy at the bottom of the valence band.  The
agreement between the two curves is very good for energies close to $E_v$. 
This means that the Fermi surface is nearly spherical at low energies as
confirmed by the calculations shown on Fig.~\ref{fig:const_surf1}. This is
due to the fact that the Fermi wavelength of the
unbound neutrons is much larger at low energies than the lattice
spacing. As a consequence the effect of Bragg diffraction is negligible.  It can be
inferred from Fig.~\ref{fig:states} that distortions of the Fermi surface
from the spherical shape happens at energies larger than 0.  The first kink
around the zero of energy is a characteristic van Hove singularity (as a
result of the vanishing of the gradient $\nabla_{\pmb{k}}\,
\varepsilon_{\alpha\pmb{k}}$ at some $\pmb{k}$ points in the expression (\ref{level_density2}), also
visible on Fig.~\ref{fig:band_spectrum}) and indicates a topological
transition of the Fermi surface.  This occurs when the Fermi sphere touches
the faces of the first Brillouin zone.  For a body centered cubic lattice,
the transition takes place when the radius of the sphere is equal to
$\sqrt{2}\pi/a$ where $a$ is the lattice spacing.  Above the first kink, the
Fermi surface becomes non spherical with the appearance of necks close to
the Brillouin zone faces as illustrated on Fig.~\ref{fig:const_surf2}. The
Fermi surface undergoes further topological changes as it crosses Bragg
planes (higher Brillouin zones) as revealed by the singularities in the
level density. The actual Fermi surface (associated with the 160 neutrons
per cell) has a very complicated shape with 11 branches (associated with the 11 bands 
which cross the Fermi level).

\begin{figure}
\begin{center}
\epsfig{file=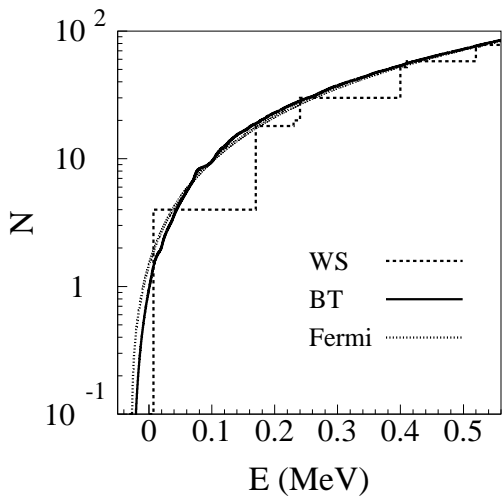,width=8.5cm}
\end{center}
\caption{Integrated state number calculated with the W-S approximation (WS,
dashed lines), band theory (BT, solid line), and free Fermi gas (doted lines)}
\label{fig:stateint}
\end{figure}

In Fig.~\ref{fig:stateint}, we show the integrated number of single 
particle levels which are below a given energy E
\begin{equation}
\label{cum_level_density}
N(E)=\int_{E_v}^E d\varepsilon\, g(\varepsilon) \, .
\end{equation}
The W-S approximation exhibits a stair steps structure due to the
discretization of the states. As already noticed, the energy levels are
discrete and highly degenerate, due to the imposed spherical symmetry
around the cluster. In contrast, in the band theory the spherical symmetry
is partly broken due to the translational symmetry of the crystal lattice. 
The energy levels thus broaden into bands, with a low residual degeneracy 
(at most 6 at each $\pmb{k}$ point for cubic crystals, counting the spin 
degeneracy~\cite{koster57}), which overlap so that the energy spectrum is continuous. 

In conclusion, for processes which involve transfered energies above the
characteristic level spacing around the Fermi energy, the differences
between the W-S approximation and the full band theory are expected to be
small.  For instance, it is typically the case for neutrino response
function~\cite{marg04} or thermal effects before the star cools completely
down. 
However at lower energies, as pertaining for instance the effective neutron mass
relevant for fluid dynamics~\cite{carter05,chamel05,chamel06}, 
the full band theory is required. The level
spacing can be roughly evaluated from the quantity $\hbar^2 /2m R_{\rm cell}^2$.
From the top to the bottom layers of the crust, the characteristic level
spacing varies from about 100~keV to 200~keV, which corresponds to temperatures of
the order of $10^9$~K. Such temperatures are found in young neutron stars
less than a few hundred years after their birth~\cite{Page06}.  The W-S
approximation is therefore well suited for describing the hot dense matter
(except for the high density layers where the spherical approximation may be
too restrictive) in young neutron stars and in the collapsing core of
massive stars during supernova explosions~\cite{bonche81}. This discussion
however does not take into account pairing effects, which are negligible in
the outermost layers of the crust considered in this work but are expected
to be important in denser and deeper layers~\cite{baldo05b}. 
It should be noted that a recent study has shown that the pairing properties 
of the unbound neutrons are strongly sensitive to the choice of boundary 
conditions in the W-S approximation, especially in the bottom layers of the crust
~\cite{baldo06,baldo06b}.

\section{Conclusion}

In this article, a comparison has been done between
the full band theory and the W-S approximation which has been
widely applied in studies of neutron star crust. The external layers of 
the inner crust at a baryon density $\rho \sim 7\times 10^{11}$ g/cm$^3$ composed 
of zirconium like clusters $^{200}$Zr have been considered. Since the bound 
nucleons are not much affected by the boundary
conditions, we have focused on the unbound neutrons. We have shown that the
ground state properties such as the unbound neutron density distribution,
are rather well reproduced by the W-S approximation, while the dynamical
properties depend on the process of interest or the energy exchanged. 
It should also be noticed that depending on the quantities of interest,
the free neutron model for the unbound neutrons could be a good first
approximation. In the future, it could be interesting to explore an
intermediate scheme which goes beyond the Wigner-Seitz approximation and
remains simpler to implement in numerical calculations than the full band
theory.

The energy spectrum is continuous in the full band theory
with no energy gaps while the W-S approximation yields a discrete spectrum
thereby overestimating neutron shell effects. The W-S approximation can
therefore be applied whenever the processes under considerations involve
energies larger than the level spacing induced by the discretization, which
in the present case is of order $\sim 100$ keV. In particular, the W-S approximation is 
 well suited for describing the hot ($T\gtrsim 10^9$ K) dense matter in the inner crust 
of young neutron stars and in the collapsing core of massive stars during supernovae 
explosions. 
However low temperature transport processes such as the effective neutron mass relating the momentum to the 
velocity and giving rise to entrainment effects, 
require a fine knowledge of the energy spectrum around the Fermi level (\textit{i.e.} 
the Fermi surface) which cannot be reproduced by the W-S approximation. 
Since the lattice spacing is predicted to decrease with increasing depth, becoming comparable to the size 
of the nuclear clusters and to the Fermi wave length of the free neutrons at the bottom of the crust, 
the validity of the W-S approximation should be carefully investigated in the denser layers of the crust, 
especially concerning pairing effects. Besides, the assumption of spherical symmetry in the W-S approximation 
is probably too restrictive near the crust-core interface where the clusters are expected to be strongly deformed~\cite{pethick95}.

\begin{acknowledgments}
N. C. gratefully acknowledges financial support from a Marie Curie Intra
European fellowship (contract number MEIF-CT-2005-024660). 
The authors acknowledge valuable discussions with N. Sandulescu, M. Pearson and 
S. Goriely during the completion of this work. 
\end{acknowledgments}


\begin{thebibliography}{99}
\bibitem{baym71} G.A. Baym, H.A. Bethe and C.J. Pethick, Nucl. Phys. {\bf A 175} (1971), 225.
\bibitem{kittel} C. Kittel, \textit{Introduction to solid state physics}, 7th edition, Wiley\&Sons (1996).
\bibitem{WS33} E.P. Wigner and F. Seitz, Phys. Rev. {\bf 43}, 804 (1933).
\bibitem{WS34} E.P. Wigner and F. Seitz, Phys. Rev. {\bf 46}, 509 (1934).
\bibitem{ne73} J.W. Negele and D. Vautherin, Nucl. Phys. {\bf A 207} (1973), 298.
\bibitem{bonche81} P. Bonche, D. Vautherin, Nucl. Phys. {\bf A 372} (1981), 496.
\bibitem{sandu04} N. Sandulescu, N. Van Giai and R.J. Liotta, Phys. Rev. {\bf C 69}, 045802 (2004).
\bibitem{sandu04b} N. Sandulescu, Phys. Rev. {\bf C 70}, 025801 (2004).
\bibitem{khan05} E. Khan, N. Sandulescu and N. Van Giai, Phys. Rev. {\bf C 71}, 042801(R) (2005).
\bibitem{vigez05} E. Vigezzi, F. Barranco, R.A. Broglia, G. Col\`o, G. Gori and F. Ramponi, Nucl. 
Phys. {\bf A 752} (2005), 600.
\bibitem{baldo05} M. Baldo, U. Lombardo, E.E. Saperstein and S.V. Tolokonnikov, Nucl. Phys. 
{\bf A 750} (2005), 409.
\bibitem{baldo06} M. Baldo, E.E. Saperstein and S.V. Tolokonnikov, Nucl.Phys. {\bf A 775} (2006), 235.
\bibitem{carter05} B. Carter, N. Chamel, P. Haensel, Nucl. Phys. {\bf A 748} (2005), 675.
\bibitem{chamel05} N. Chamel, Nucl.Phys. {\bf A 747} (2005), 109.
\bibitem{chamel06} N. Chamel, Nucl.Phys. {\bf A 773} (2006), 263.
\bibitem{pethick95} C.J. Pethick and D.G. Ravenhall, Ann. Rev. Nucl. Part. Sci. {\bf 45} (1995), 429.
\bibitem{baldo05b} M. Baldo, E.E. Saperstein and S.V. Tolokonnikov, Nucl. Phys. {\bf A 749} (2005), 42.
\bibitem{magier03} P. Magierski, A. Bulgac, P.-H. Heenen, Nucl. Phys. {\bf A 719} (2003), 217.
\bibitem{chabanat98} E. Chabanat, P. Bonche, P. Haensel, J. Meyer, R. Schaeffer, Nucl. Phys. 
{\bf A 635} (1998), 231 ; Erratum, Nucl. Phys. {\bf A 643} (1998), 441.
\bibitem{koster57} G.F. Koster, Solid State Physics {\bf 5}, edited by F. Seitz, D. Turnbull, Academic Press, 
New York (1957), 173.
\bibitem{baldo06b} M. Baldo, E.E. Saperstein, S.V. Tolokonnikov, arxiv preprint nucl-th/0609031.
\bibitem{marg04} J. Margueron, J. Navarro, and P. Blottiau, Phys. Rev. {\bf C 70}, 028801 (2004).
\bibitem{Page06} D. Page, U. Geppert, F. Weber, Nucl. Phys. {\bf A 777} (2006), 497.
\end{thebibliography}
\end{document}